\documentclass[letterpaper, 10 pt, conference]{ieeeconf}

\IEEEoverridecommandlockouts                              
                                                          
\usepackage[utf8]{inputenc}
\usepackage{adjustbox}
\usepackage{amsmath,amssymb,amsfonts}
\usepackage{algorithmic}
\usepackage{multirow}
\usepackage{graphicx}
\usepackage{textcomp}
\usepackage{xcolor}
\usepackage{hyperref}
\usepackage{diagbox}
\usepackage[table]{xcolor}
\usepackage{colortbl}
\usepackage{adjustbox}
\usepackage{diagbox}
\usepackage{flushend}

\usepackage{tikz}
\newcommand{\copyrightnotice}{
  \begin{tikzpicture}[remember picture,overlay]
    \node[anchor=south,yshift=10pt] at (current page.south) {
      \parbox{\textwidth}{\footnotesize © 2026 IEEE. Personal use of this material is permitted. Permission from IEEE must be obtained for all other uses, in any current or future media, including reprinting/republishing this material for advertising or promotional purposes, creating new collective works, for resale or redistribution to servers or lists, or reuse of any copyrighted component of this work in other works.}
    };
  \end{tikzpicture}
}

\definecolor{low}{RGB}{222,235,247}
\definecolor{mid}{RGB}{158,202,225}
\definecolor{high}{RGB}{49,130,189}

\newcommand{\shadecell}[1]{%
  \ifnum#1>5 \cellcolor{high}#1
  \else\ifnum#1>3 \cellcolor{mid}#1
  \else\ifnum#1>0 \cellcolor{low}#1
  \else #1 \fi\fi\fi
}
\hypersetup{colorlinks=true,}
\newcommand{\shadecellperc}[2]{%
  \ifnum#1>5 \cellcolor{high}#2
  \else\ifnum#1>3 \cellcolor{mid}#2
  \else\ifnum#1>0 \cellcolor{low}#2
  \else #2 \fi\fi\fi
}

\newcommand{\textcomment}[1]{} 

\usepackage{xcolor}
\definecolor{jenish_colour}{rgb}{.0,.6,.05}
\definecolor{samarth_colour}{rgb}{0,0,1}
\definecolor{vineet_colour}{rgb}{0,0.35,0}

\usepackage[sorting=none]{biblatex} 
\title{\LARGE \bf
Towards Real-World Identification of Fatigued Muscle Groups via Musculoskeletal Simulation
}
\author{Jenishkumar Chauhan, Samarth Brahmbhatt, and Vineet Vashista 
\thanks{Jenishkumar Chauhan is with the Human-Centered Robotics Lab at IIT Gandhinagar, Gujarat, India as a PhD student.
        {\tt\small jenishkumar.chauhan@iitgn.ac.in}}%
\thanks{Samarth Brahmbhatt is an independent researcher from Mountain View, California, USA.
        {\tt\small samarth.robo@gmail.com}}%
\thanks{Vineet Vashista$^{*}$ is with the Human-Centered Robotics Lab at IIT Gandhinagar, Gujarat, India as an associate professor. 
        {\tt\small vineet.vashista@iitgn.ac.in}}%
\thanks{$^{*}$corresponding author}
}

\begin{document}

\maketitle
\copyrightnotice
%
%
%
%
\begin{abstract}
Contactless diagnosis of musculoskeletal disorders can potentially improve population health as well as robot behaviours in collaborative settings. However, current diagnosis methods require an in-person physical examination in which a trained physician senses, through contact, the force applied by various muscles. Simulation tools exist, but their use for diagnosis with real data is under-explored. In this paper, we propose an algorithm for identifying which upper-limb muscle group is fatigued. Our algorithm compares the real-world free-space motion of the subject with that of a simulated musculoskeletal model, and is therefore \emph{contactless}: preventing the need for invasive sensing or in-person assessment. Our algorithm simulates various fatigue conditions using a physics-based musculoskeletal model and extracts diagnostic motion features from both real and simulated data, which are compared for diagnosis. Experimental results on real data demonstrate that the proposed method can reliably distinguish between multiple muscle-groups of fatigue. Additionally, through comprehensive performance comparisons, we show how recent advanced musculoskeletal simulators can be properly configured to address the sim-to-real gap in the context of the fatigue diagnosis task. Our approach can potentially spur further research in remote and automated diagnosis, significantly lowering the barrier to large-scale and early detection.
\end{abstract}


\begin{figure*}
\centerline{\includegraphics[width=\linewidth]{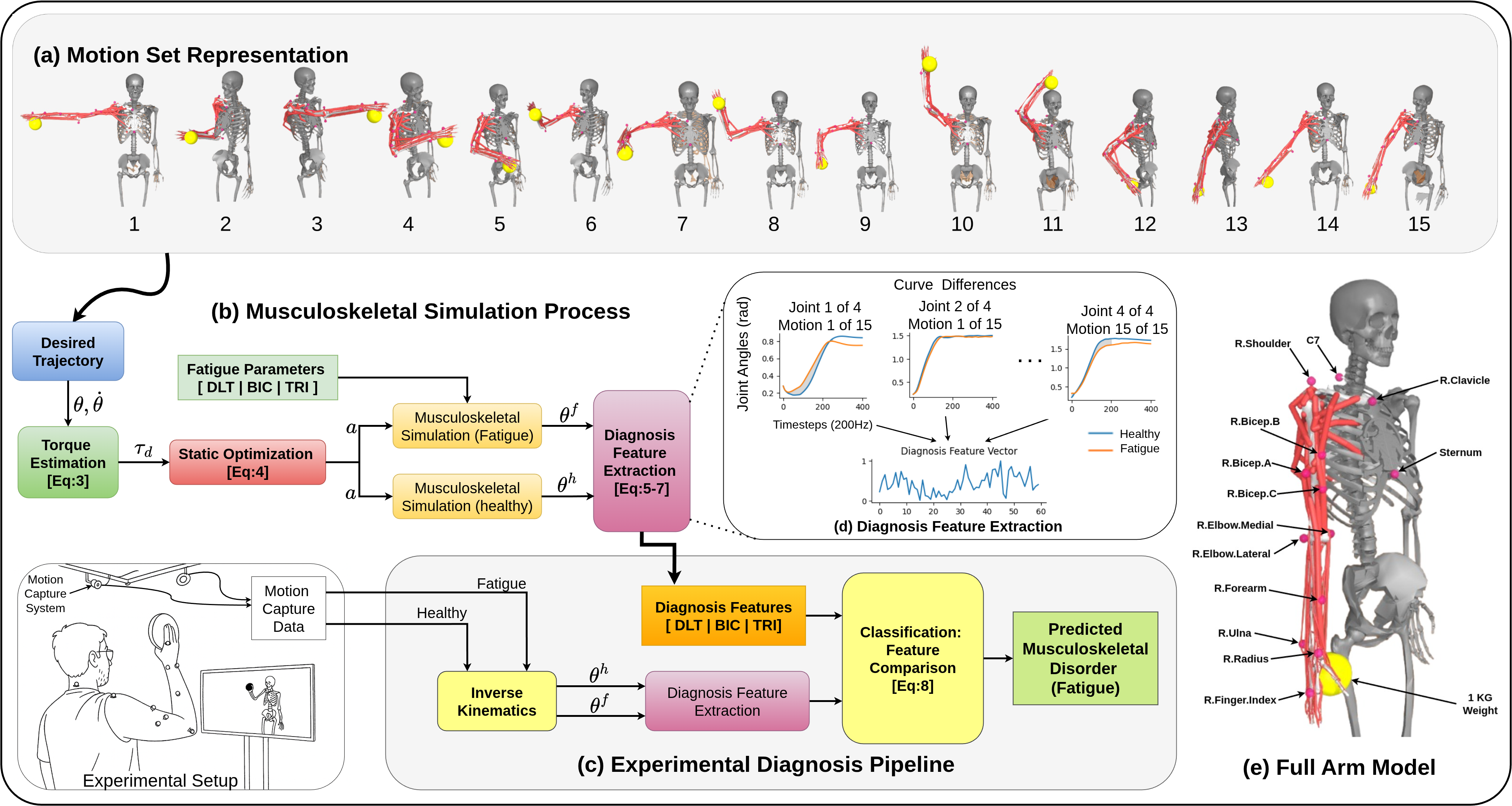}}
\caption{ \textbf{Overview of the simulation-driven diagnosis framework for identifying fatigue in upper-limb muscle groups.} \textbf{(a) Motion Set Representation:} A set of 15 representative shoulder–elbow motions performed with a 1 kg weight was used as the basis for both simulation and experimental trials.  \textbf{(b) Musculoskeletal Simulation Process:} Desired joint trajectories 
(\(\theta, \dot{\theta}\)) are converted into desired torques via torque estimation [Eq. (3)], followed by static optimization [Eq. (4)] to map torques into muscle activations. Simulations were executed under both healthy and fatigued conditions by modulating muscle-specific parameters (DLT: deltoid, BIC: biceps, TRI: triceps). The simulated joint kinematics (\(\theta^h, \theta^f\)) were used for diagnosis feature extraction [Eqs. (5–7)].  \textbf{(c) Experimental Diagnosis Pipeline:} Human participants reproduced the same motion set while holding a 1 kg weight. Marker-based motion capture data were processed through inverse kinematics to obtain physiological joint angles (3 shoulder and 1 elbow). Diagnosis feature vectors were extracted from both simulation and experimental data, and classification was performed by feature comparison [Eq. (8)] to predict the most likely fatigued muscle group.  \textbf{(d) Diagnosis Feature Extraction:} Example plots show healthy (blue) vs fatigued (orange) joint trajectories across different motions. The deviation between healthy and fatigued simulations was quantified into a normalized diagnosis feature vector (60 dimensions = 15 motions × 4 joints).  \textbf{(e) Full Arm Model:} Musculoskeletal model adapted from MyoSuite, comprising 38 joints and 63 muscles. Markers (pink) were attached on anatomical landmarks (e.g., R.Shoulder, R.Elbow, R.Biceps) to align with the experimental motion capture system.} 

\label{fig:main}
\end{figure*}

\section{Introduction and Related Work}

Musculoskeletal disorders (MSDs)~\cite{gallagher2020musculoskeletal} are a group of conditions that affect the muscles, bones, tendons, ligaments, and other parts of the musculoskeletal system. These disorders may arise from a variety of causes, including injury, overuse, repetitive strain, poor posture, and aging. MSDs are among the leading global causes of disability ~\cite{gill2023global}, characterized by pain, functional impairment, and reduced mobility due to conditions affecting the muscles, bones, joints, or connective tissues. They can affect various regions of the body, leading to stiffness, swelling, and limitations in movement during daily or task-specific activities. Common examples include muscle fatigue, rotator cuff injuries, tendinopathy, osteoarthritis, frozen shoulder, and sarcopenia. Fatigue refers to a reduction in muscle force-generating capacity due to prolonged or repetitive activity. It manifests as decreased joint stability, altered coordination, and compensatory movements, which can be captured through deviations in joint-space trajectories \cite{enoka2008muscle, gandevia2001spinal}. 

Robots collaborating with humans on a shared task can benefit from an accurate assessment of the fatigue location and severity of the human partners. This information can modulate the robot's action speed, for example. However, current diagnostic methods for MSDs primarily rely on an in-person clinical physical exam, such as range of motion (ROM) tests and manual muscle testing~\cite{wilson1990musculoskeletal}. This motivates the need for sensor-based fatigue diagnosis methods that can operate without physical contact with the subject.


Medical imaging (e.g., MRI, X-ray, ultrasound)~\cite{patel2012review, leardini2005human, luomajoki2008movement} is useful for structural assessment but often requires bulky equipment and lacks the resolution to detect subtle, functional, or early-stage impairments. Surface electromyography (sEMG) is also employed for muscle activation analysis~\cite{pantelopoulos2009survey}, but it is limited by noise, skin-electrode variability, and poor suitability for continuous or naturalistic movement assessments~\cite{camomilla2018trends}.

On the other hand, algorithms like~\cite{mehta2017vnect, mathis2018deeplabcut} can track upper-limb kinematics from a video stream alone or in conjunction with wearable sensors, and have been used for biomedical motion analysis \cite{nakano2020evaluation, robert2020validation, mundermann2006evolution} and even fatigue assessment~\cite{yu2019automatic}. Our proposed algorithm operates on similar kinematic data, but in contrast to broad fatigue level assessment, it predicts which muscle group is fatigued. We source our experimental data in this paper from a 3D motion capture setup.


Another alternative is to fit musculoskeletal models to experimentally captured data. Recently works like \cite{roupa2022modeling, killen2020silico} have demonstrated how models can capture patient-specific motion characteristics and support diagnostic decision-making. Such modeling-based diagnostic frameworks have also been validated in clinical studies \cite{van2016neuro}. However, these methods often require bespoke instrumentation of the affected body parts. In this work, we leverage a general-purpose musculoskeletal simulator called MyoSuite~\cite{caggiano2022myosuite}. The upper-limb kinematics data our algorithm uses can be sourced from well-known systems like 3D motion capture or camera-based 3D pose estimation, which do not require as much specialized knowledge to set up.



To summarize, we make the following contributions in this paper:
\begin{itemize}
    \item \textbf{Simulator setup}: We show through diagnosis performance evaluations how a musculoskeletal simulator can be configured to generate motion data for accurate real-world fatigue identification.
    \item \textbf{Diagnostic features}: We propose novel features describing the difference of motion between healthy and fatigued states, which can be compared meaningfully across simulation and reality.
    \item \textbf{Experiments}: We show that our proposed algorithm can reliably identify the fatigued muscle group through experiments on real participant data.
\end{itemize}

All the data used in this paper and corresponding diagnosis code will be released publicly.
\section{Method}

This study aims to establish a simulation-driven framework for the contactless diagnosis and monitoring of fatigue in muscle groups of the upper limb and torso. By integrating experimental motion capture with musculoskeletal simulations, it provides a means to detect and quantify kinematic deviations caused by localized impairments, such as fatigue. This approach offers potential applications in early screening, rehabilitation feedback, and assistive device design, enabling subject-specific and non-invasive assessment of human motion.

The overall simulation-driven diagnosis can be conceptualized in two levels. The lower-level control focuses on simulating musculoskeletal motion under both healthy and impaired conditions, driven by kinematic data extracted from participants through motion capture. At the higher level, the framework focuses on extracting motion diagnosis features from these simulations and leveraging them to classify experimental motion capture data into impairment-specific categories, such as fatigue in particular muscle groups such as the deltoid (referred to as DLT), which includes anterior, lateral, and posterior deltoid muscles for shoulder elevation and rotation, biceps (BIC) comprising the long and short heads responsible for elbow flexion and forearm supination, and triceps (TRI) including the long, lateral, and medial heads that extend the elbow.

\subsection{Musculoskeletal Model and Simulation Environment}
The musculoskeletal model employed in this study is implemented using the the MyoSuite framework~\cite{wang2022myosim}, which uses the MuJoCo physics engine~\cite{todorov2012mujoco} internally. We use the full-arm model, which comprises 38 joints and 63 muscles, providing detailed anatomical representation of the upper limb. MyoSuite provides XML-based models such as MyoArm, which specify detailed skeletal geometry, joint definitions, and muscle-tendon routing based on anatomical fidelity, making it suitable for realistic simulation of arm movements.


We disabled the muscle actuators associated with the forearm, wrist, and palm in order to focus the study on the upper arm and torso. Note that theoretically, our proposed algorithm is agnostic to the muscle group being chosen for diagnosis.

The forward simulation process in MuJoCo consists of several computational layers, beginning with normalized actuator control signals, which are smoothed into effective muscle activations by MuJoCo activation dynamics. These activations generate muscle forces via gain and bias parameters: the gain represents active force modulated by activation, while the bias accounts for passive contributions from force–length–velocity relationships \cite{millard2013flexing, caggiano2022myosuite}. The resulting muscle forces act on joints and are mapped into torques through moment arms, ultimately producing joint kinematics (angles, velocities, and accelerations) that match musculoskeletal dynamics.

Joint torque is expressed as a non-linear function of muscle activation, tendon length, and contraction velocity:
\begin{align}
F_{\text{muscle}} &= {\text{gain}} \cdot {\text{activations}}  + {\text{bias}} \label{eq:muscle_force} \\  
\tau &= M_A^T \times F_{\text{muscle}}  \label{eq:joint_torques}
\end{align}
where total muscle force \(F_{\text{muscle}}\) is the sum of active and passive components. The moment arm matrix \({M_A^T}\), derived from the model configuration, maps muscle forces to joint torques.

\subsection{Motion Generation and Control Strategy} \label{sec:refernce_trajectory}
As part of the control pipeline, the desired trajectories were used to guide the simulation. These were derived from human arm motions recorded via a Vicon marker-based motion capture system and represented as matrices of joint angles for the shoulder and elbow. A set of 15 representative trajectories was selected (Fig.~\ref{fig:main}(a)) to evaluate performance across varied musculoskeletal conditions. These motions included coordinated shoulder and elbow activities, chosen to reflect common postures and movement primitives observed in daily activities. Each trajectory served as a reference for both healthy and impaired simulations, providing a standardized set of inputs for comparison. A supplementary video illustrates these motion examples in detail.

\subsubsection{Inverse Kinematics}
To ensure consistency between the simulated and experimentally recorded motion data, the same motion capture marker template used in experiments (13 markers, as shown in Fig.~\ref{fig:main}(e)) was applied to the simulation model. From these markers, physiological joint angles (3 shoulder angles and 1 elbow angle) were computed and subsequently used in the motion diagnosis analysis.
In addition to these physiological joint angles, the simulation environment operates with joint angles defined according to the MyoSuite model configuration. These simulation-specific joint angles were estimated using a least-squares solver, which iteratively explored forward kinematic combinations in the simulator to ensure that the simulated physiological joint angles matched the experimentally derived values as closely as possible.

\subsubsection{Joint Torque Computation}
Desired torques were obtained using a proportional–derivative (PD) controller that minimized the difference between current and target states:
\begin{align} \label{eq:PD_control}
&\tau_{\text{desired}} = K_p (\theta_{\text{target}} - \theta_{\text{current}}) + K_d (\dot{\theta}_{\text{target}} - \dot{\theta}_{\text{current}}) 
\end{align}
where $\tau_{\text{desired}}$ is the desired joint torque, $\theta$ is the joint angle, and $\dot{\theta}$ is the joint angular velocity. $K_p$ and $K_d$ are the proportional and derivative gains, tuned heuristically to minimize response time and oscillations in the motion. 

\subsubsection{Static optimization - Muscle Activation Estimation}
Desired torques were then mapped to muscle activations (\(a\)) through static optimization, formulated as a quadratic programming (QP) problem:
\begin{align}
\min_{a} \sum_i a_i^2 \quad \text{s/t:} \quad \mathcal{F}(a, l, v) = \tau_{\text{desired}}, \quad 0 \leq a_i \leq 1
\end{align}
\(a_i\) is muscle activation, \(\mathcal{F}(a, l, v)\) represents torque generation through activation dynamics, force–length–velocity relationships~\cite{millard2013flexing}, and joint moment arms. \({l}\), \({v}\) denote the current muscle length and velocity respectively.

This optimization minimizes the sum of squared activations while ensuring torque consistency, subject to anatomical and physiological constraints, following classical static optimization approaches in biomechanics \cite{crowninshield1981physiologically}. MuJoCo's internal solvers handled this mapping by simulating the muscle activation to muscle forces to torque transformation based on anatomical routing and muscle dynamics. The target trajectories, represented through joint angles in the shoulder and elbow, were defined based on human motion capture data as described above. 
\begin{figure} [ht!]
\centerline{\includegraphics[width=1\linewidth]{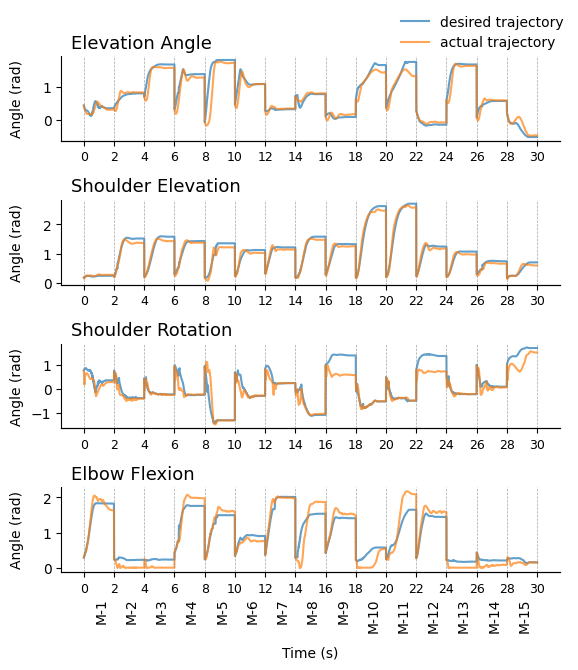}}
\vspace{-10pt}
\caption{\textbf{Tracking of 15 representative upper-limb motions in the musculoskeletal simulation.} Desired joint trajectories (blue) derived from motion capture were compared with simulated joint responses (orange) across four anatomical angles: elevation angle, shoulder elevation, shoulder rotation, and elbow flexion. Each motion segment (M-1 to M-15) lasted approximately 2 seconds, separated by vertical dashed lines. The close alignment between desired and simulated trajectories demonstrates the model’s ability to reproduce physiological kinematics under healthy conditions, forming the baseline for subsequent fatigue-induced comparisons.}
\label{fig:LL}
\end{figure}
This pipeline ensures biologically plausible muscle-driven motion generation aligned with human anatomical and mechanical constraints. The resulting muscle activations are used to drive the simulated model, ensuring that the generated motion accurately reflects the underlying biomechanics.

\subsubsection{Simulation stepping durations}
By default, the MyoSuite environment steps every 20 ms and the internal MuJoCo physics engine steps every 2 ms. Our motion capture trajectories are captured at 200 Hz. It is crucial for accurate tracking that we decrease the stepping duration of MyoSuite and MuJoCo to 5 ms and 0.5 ms respectively to match the motion capture trajectories, rather than sub-sampling the trajectories. Section~\ref{sec:ablation} shows the impact of this on the overall diagnosis accuracy.

The performance of the healthy musculoskeletal model in simulation is illustrated in Fig.~\ref{fig:LL}, where the desired trajectories derived from 15 experimental motions are compared against the simulated joint responses for four representative joints, including three shoulder angles and one elbow angle. Each motion lasted for 2 seconds, after which the simulation was reset to its initial state before the subsequent motion. The close alignment between the desired (blue) and simulated (orange) trajectories across all motions demonstrates the model’s ability to accurately reproduce experimental kinematics under healthy conditions. This provides a validated baseline for subsequent comparisons with impaired simulations, where deviations in trajectory tracking are expected to reflect the localized effects of musculoskeletal disorders.

\subsection{Modeling of Musculoskeletal Disorders}
Various musculoskeletal disorders, such as sarcopenia and neuromuscular fatigue, can be modeled by modifying muscle properties \cite{caggiano2022myosuite}, altering the anatomy of the musculoskeletal arrangements, and introducing specific behavioral dynamics in MyoSuite and MuJoCo. In the case of fatigue, localized impairments were simulated by reducing the force production capacity of the targeted muscles. This was implemented as scaling their gains down by a `fatigue factor', which directly influences the ability of the muscle to generate force. Importantly, only muscles belonging to the specific group were altered, thereby isolating the effects of group-specific degradation while preserving the behavior of other muscles in the arm.

This study aimed at three primary upper arm muscles, the biceps (BIC), the triceps (TRI), and the deltoids (DLT)~\cite{corcos2002fatigue}, to assess the localized impact of impairments. It is important to set the fatigue factor values such that the motion characteristics in simulation are affected similarly to the real participants. Because it is difficult to measure muscle gains directly in the body, we measured the relevant muscle activations in the healthy and fatigued conditions while a participant was performing a pre-defined motion over a one minute duration. Muscle activation values are readily available in the simulation, while for the real participant we measured them with surface EMG electrodes \cite{zhou2011evaluation, rota2014influence}. We varied the fatigue factor in simulation such that the ratio of healthy and fatigued RMS muscle activations matched as closely as possible between simulation and reality. Note that an exact match is not possible because we used the same fatigue factor value across all muscles in the muscle group. Table \ref{tab:muscle_group_values} shows the RMS muscle activation ratios for the real participant and simulation, followed by the tuned fatigue factor values.


\begin{table}[ht!]
\centering \small 
\renewcommand{\arraystretch}{1.25}
\begin{adjustbox}{width=1\columnwidth}
\begin{tabular}{|c|c|c|c|c|c|c|c|c|}
\hline
\multirow{2}{*}{\textbf{Source}} & \multicolumn{3}{c|}{\textbf{BIC Group}} & \multicolumn{3}{c|}{\textbf{DLT Group}} & \multicolumn{2}{c|}{\textbf{TRI Group}} \\
\cline{2-9}
& \textbf{BIC long} & \textbf{BIC short} & \textbf{BRA} & \textbf{ANT DLT} & \textbf{LAT DLT} & \textbf{POST DLT} & \textbf{TRI long} & \textbf{TRI LAT} \\
\hline
real participant & 0.782 & 0.834 & 0.767 & 0.651 & 0.842 & 1.302 & 0.561 & 0.714 \\
simulation & 0.838 & 0.838 & 0.838 & 0.772 & 0.904 & 0.982 & 0.652 & 0.652 \\ \hline
fatigue factor & 0.720 & 0.720 & 0.720 & 0.480 & 0.480 & 0.480 & 0.680 & 0.680 \\
\hline
\end{tabular}
\end{adjustbox}
\vspace{4pt}
\caption{Ratio of RMS muscle activations across real participant, simulation, and the tuned fatigue factors. \\A fatigue factor of 1 represents the healthy condition, while lower values correspond to higher muscle fatigue.}
\label{tab:muscle_group_values}
\end{table}

\subsection{Kinematic Comparison and Diagnosis Feature Vector}

\subsubsection{Joint Angle Mapping} 
Each simulated motion generates 38 joint angles as defined by the MyoSuite full-arm model configuration. Notably, the shoulder joint alone accounts for 14 of these values, which correspond to a detailed internal representation of the 3 anatomical degrees of freedom. For meaningful comparison with experimental motion capture data, simplified anatomical inverse kinematics (IK) template is defined using motion-capture marker-based reconstruction. This reduced model representation includes only 4 joint angles, 3 for the shoulder and 1 for the elbow, which is consistent with standard biomechanical conventions used for analyzing upper-limb motion in experimental settings.

\subsubsection{Diagnosis Feature} \label{sec:diagnostic_features}

To quantify the deviation between healthy and fatigued motion trajectories, a joint-wise mean squared error (MSE) was computed for each motion task. For the \(k\)-th motion and the \(j\)-th joint, the MSE is given by:
\begin{align} \label{eq:curve_error}
E_k^{(j)} &= \frac{1}{T_k} \sum_{t=1}^{T_k} \left( {\theta_j^h}^{(k)}(t) - {\theta_j^i}^{(k)}(t) \right)^2
\end{align}
where \({\theta_j^h}^{(k)}(t)\) and \({\theta_j^i}^{(k)}(t)\) represent the \(j^{th}\) joint angle trajectories under healthy and impaired conditions such as fatigue, respectively, and \(T_k\) is the number of time steps in the \(k\)-th motion.
To ensure comparability across joints, the error values were normalized using Min-Max normalization.
\begin{align} \label{eq:normalization}
\hat{E}_k^{(j)} = \frac{E_k^{(j)} - \min_k E_k^{(j)}}{\max_k E_k^{(j)} - \min_k E_k^{(j)} + \varepsilon}
\end{align}
All normalized errors were assembled into a matrix \(\mathbf{E} \in \mathbb{R}^{J \times K}\), where \(J\) is the number of joints and \(K\) is the number of motions. Finally, this matrix was vectorized column-wise to obtain the diagnosis feature vector:
\begin{align} \label{eq:diag_vector}
\mathbf{f}_{\text{diag}} = \mathrm{vec}(\mathbf{E}) \in \mathbb{R}^{JK}
\end{align}
This feature vector \( \mathbf{f}_{\text{diag}} \) captures joint-wise kinematic deviations across all motions and is used for subsequent classification and diagnostic tasks. In this work, it results in a 60-dimensional vector per subject, corresponding to 15 motions across 4 joints.

\subsection{Contactless Diagnosis}
To evaluate the proposed simulation-based approach, Diagnosis Feature Vectors generated from simulated disorder cases are compared against those extracted from experimental trials involving healthy and fatigued subjects. A close match between the simulated and experimental feature vectors indicates that the simulated condition closely reproduces real-world motion deviations. This comparison forms the basis of a contactless diagnostic framework, enabling scalable and non-invasive detection of musculoskeletal impairments by matching observed motion to pre-simulated fatigue conditions.
The predicted disorder class \(\hat{c}\) is identified by selecting the simulation whose diagnosis feature vector is closest (in root-mean-square sense) to the experimental feature vector:
\begin{align}
&\hat{c} = \arg\min_{c} \sqrt{ \left\| \mathbf{f}_{\text{diag}}^{\text{exp}} - \mathbf{f}_{\text{diag}}^{sim(c)} \right\|^2}
\end{align}

here, \(\mathbf{f}_{\text{diag}}^{\text{exp}}\) is the Diagnosis Feature Vector obtained from the experimental motion data of a participant, and \(\mathbf{f}_{\text{diag}}^{sim(c)}\) is the feature vector from the simulation of disorder class \(\hat{c}\). The disorder class with the smallest discrepancy is considered the most likely underlying impairment for the observed motion.


\section{Experiments}
To validate the fatigue motion behavior observed in simulation, a set of 15 representative arm motions was reproduced experimentally by human participants, both under healthy conditions and after inducing fatigue in specific muscle groups. Motion data were recorded using a Vicon motion capture system and analyzed using the same methodology applied in the simulation.

\subsection{Experiment protocol}
A total of 21 fatigue trials were conducted, with seven trials for each fatigue case (DLT, BIC, and TRI). These trials were performed by nine healthy participants (age: 22 ± 2 years, gender: male, BMI: 23.95 ± 4.01), free from musculoskeletal or neurological disorders. Some participants completed all three fatigue conditions, while others completed only two. Thirteen reflective markers were placed on key anatomical landmarks of the upper limb to calculate physiological joint angles, similar to the simulation description as per Fig-1(e). The predefined sequence of motions shown to the participant as a video of a musculoskeletal model performing 15 motion sequences as described in Figure-1(a), attached in supplementary material.
Throughout the experiment, participants held a constant 1 kg weight in their hand to increase task difficulty. Each subject performed the motion sequence under two conditions: (i) healthy baseline and (ii) after targeted fatiguing exercises for the biceps (BIC), triceps (TRI), or deltoid (DLT). Fatigue was induced through repetitive exercises until maximum voluntary exhaustion, or until at least 100 repetitions were completed. The captured kinematic data were processed to obtain joint angles, which were then compared against simulation benchmarks using the Diagnosis Feature Vector. 

\subsection{Fatigued Muscle-group Identification}
Four joint angles (three at the shoulder and one at the elbow) were analyzed to identify deviations between healthy and fatigued states. Diagnosis Feature Vectors, as described in Eqs. \ref{eq:curve_error}, \ref{eq:normalization}, \ref{eq:diag_vector}, were computed for each trial by comparing baseline (healthy) data against fatigued motion data of the same participant.

For each trial, the root mean square difference (RMSD) was computed between the experimental Diagnosis Feature Vector and the simulation-derived feature vectors for each fatigue condition. These RMSD values shows the similarity of experimental diagnosis feature vector to the corresponding fatigue diagnosis feature vector. The closest match, lowest RMSD value is considered as probable case of fatigue, identifying the specific muscle group (biceps, triceps, or deltoid) that was fatigued during the corresponding fatigue-induced sequence.

\begin{table}[ht!] 
\centering \small 
\renewcommand{\arraystretch}{1.4} \scriptsize
\begin{adjustbox}{width=1\columnwidth}
\begin{tabular}{|c|ccc|}
\hline
\diagbox{True}{Pred} & DLT group & BIC group & TRI group \\
\hline
Experimental DLT group & \shadecell{5} & \shadecell{0} & \shadecell{2} \\
Experimental BIC group & \shadecell{1} & \shadecell{6} & \shadecell{0} \\
Experimental TRI group & \shadecell{2} & \shadecell{1} & \shadecell{4} \\
\hline
\end{tabular}
\end{adjustbox}
\vspace{3pt}
\caption{Confusion matrix for experimental diagnosis accuracy.
Classification outcomes for 21 trials (7 per fatigue group) using simulation-derived diagnosis features compared against experimental data.}
\label{tab:confusion_matrix}
\end{table}

Table \ref{tab:confusion_matrix} summarizes the confusion matrix comparing simulation benchmarks with experimental data. Across 21 fatigue trials (7 each for Deltoid, Biceps, and Triceps), the Diagnosis Feature approach yielded an overall classification precision of 71.43\%.
\begin{figure}[htbp]
\centerline{\includegraphics[width=1\linewidth]{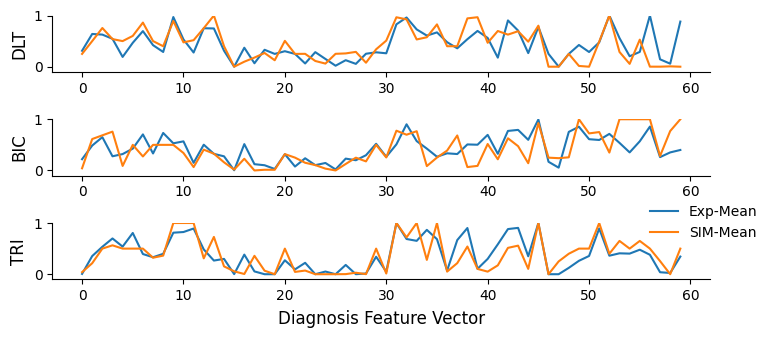}}
\caption{\textbf{Comparison of simulation- and experiment-derived diagnosis feature vectors across fatigue groups.} Mean normalized diagnosis feature vectors (60 dimensions: 15 motions × 4 joints) are shown for the deltoid (DLT), biceps (BIC), and triceps (TRI) fatigue cases. The experimental curves (blue) represent the \textbf{mean across all participants}, while the simulation curves (orange) denote the corresponding simulated fatigue conditions. The close alignment between experimental means and simulations indicates that the simulated fatigue models capture the primary kinematic deviations observed in participants, supporting the use of simulation-driven features for motion-based fatigue diagnosis.}
\vspace{-8pt}
\label{fig:mean_diagnosis_feature} 
\end{figure}

Figure \ref{fig:mean_diagnosis_feature} presents the overall trend of experimental and simulation-derived Diagnosis Feature values. It illustrates how these values vary across different fatigue cases and how closely the simulated fatigue conditions align with the corresponding experimental mean values. These results indicate that Diagnosis Feature-based diagnosis achieved an overall macro precision of 71.43\% across the study population, underscoring the potential of simulation-driven, contactless diagnosis of muscular fatigue.

\subsection{Ablation Studies and Statistical Validation} \label{sec:ablation}

\subsubsection{Fatigue factor tuning}
This ablation study varied the fatigue scaling factors across muscle groups (BIC, DLT, TRI). For each configuration, diagnosis feature vectors were computed and matched against experimental data to observe changes in classification outcomes. Results shown in Table~\ref{tab:ablation_confusions_FR} indicate that very subtle scaling (e.g., 0.9) caused overfitting to a single muscle group, while extreme fatigue scaling (e.g., 0.1) produced more balanced results. A manually tuned non-uniform configuration (e.g., 0.48–0.72–0.68) achieved the best separation, underscoring the importance of realistic, muscle-specific simulation modeling.

\begin{table}[ht!]
\centering
\renewcommand{\arraystretch}{1.2}
\setlength{\tabcolsep}{4pt}
\begin{adjustbox}{width=1.0\columnwidth}
\begin{tabular}{ccc}
\textbf{(a) Fatigue Rate: 0.9, 0.9, 0.9} & \textbf{(b) Fatigue Rate: 0.75, 0.75, 0.75} & \textbf{(c) Fatigue Rate: 0.48, 0.72, 0.68} \\
\begin{tabular}{|c|ccc|}
\hline
\diagbox{True}{Pred} & DLT & BIC & TRI \\
\hline
DLT & \shadecell{4} & \shadecell{3} & \shadecell{0} \\
BIC & \shadecell{0} & \shadecell{7} & \shadecell{0} \\
TRI & \shadecell{2} & \shadecell{5} & \shadecell{0} \\
\hline
\end{tabular}
&
\begin{tabular}{|c|ccc|}
\hline
\diagbox{True}{Pred} & DLT & BIC & TRI \\
\hline
DLT & \shadecell{4} & \shadecell{3} & \shadecell{0} \\
BIC & \shadecell{0} & \shadecell{7} & \shadecell{0} \\
TRI & \shadecell{2} & \shadecell{5} & \shadecell{0} \\
\hline
\end{tabular}
&
\begin{tabular}{|c|ccc|}
\hline
\diagbox{True}{Pred} & DLT & BIC & TRI \\
\hline
DLT & \shadecell{5} & \shadecell{0} & \shadecell{2} \\
BIC & \shadecell{1} & \shadecell{6} & \shadecell{0} \\
TRI & \shadecell{2} & \shadecell{1} & \shadecell{4} \\
\hline
\end{tabular}
\\[12pt]
\textbf{(d) Fatigue Rate: 0.5, 0.5, 0.5} & \textbf{(e) Fatigue Rate: 0.25, 0.25, 0.25} & \textbf{(f) Fatigue Rate: 0.1, 0.1, 0.1} \\
\begin{tabular}{|c|ccc|}
\hline
\diagbox{True}{Pred} & DLT & BIC & TRI \\
\hline
DLT & \shadecell{1} & \shadecell{5} & \shadecell{1} \\
BIC & \shadecell{0} & \shadecell{5} & \shadecell{2} \\
TRI & \shadecell{1} & \shadecell{3} & \shadecell{3} \\
\hline
\end{tabular}
&
\begin{tabular}{|c|ccc|}
\hline
\diagbox{True}{Pred} & DLT & BIC & TRI \\
\hline
DLT & \shadecell{3} & \shadecell{4} & \shadecell{0} \\
BIC & \shadecell{1} & \shadecell{6} & \shadecell{0} \\
TRI & \shadecell{3} & \shadecell{2} & \shadecell{2} \\
\hline
\end{tabular}
&
\begin{tabular}{|c|ccc|}
\hline
\diagbox{True}{Pred} & DLT & BIC & TRI \\
\hline
DLT & \shadecell{6} & \shadecell{1} & \shadecell{0} \\
BIC & \shadecell{0} & \shadecell{5} & \shadecell{2} \\
TRI & \shadecell{2} & \shadecell{1} & \shadecell{4} \\
\hline
\end{tabular}
\end{tabular}
\end{adjustbox}
\vspace{4pt}
\caption{Impact of fatigue factor tuning on the overall diagnosis performance.}
\label{tab:ablation_confusions_FR}
\end{table} 

\subsubsection{Statistical significance of diagnostic features}
In the second analysis, pairwise t-tests were applied across the 60-dimensional diagnosis feature space to compare fatigue conditions. High inter-subject variability meant many features did not reach statistical significance (p \(\leq\) 0.05). To reduce noise, features were progressively filtered using increasing p-value thresholds (e.g., $p \leq 0.1$, 0.25, 0.5), and classification performance was recomputed. Using only the most discriminative features (2–11 features at $p \leq 0.1$–0.25) improved accuracy, especially for BIC and TRI fatigue. In contrast, including all features ($p \leq 1.0$) diluted separability, as shown in Table \ref{tab:confusion_pval_thresh}.

\begin{table}[ht!]
\centering
\renewcommand{\arraystretch}{1.2}
\setlength{\tabcolsep}{4pt}
\begin{adjustbox}{width=0.67\columnwidth}
\begin{tabular}{cc}
\textbf{(a) p-value $\leq$ 0.10 (2 features)} & \textbf{(b) p-value $\leq$ 0.25 (11 features)} \\
\begin{tabular}{|c|ccc|}
\hline
\diagbox{True}{Pred} & DLT & BIC & TRI \\
\hline
DLT & \shadecell{5} & \shadecell{1} & \shadecell{1} \\
BIC & \shadecell{0} & \shadecell{7} & \shadecell{0} \\
TRI & \shadecell{4} & \shadecell{0} & \shadecell{3} \\
\hline
\end{tabular}
&
\begin{tabular}{|c|ccc|}
\hline
\diagbox{True}{Pred} & DLT & BIC & TRI \\
\hline
DLT & \shadecell{3} & \shadecell{2} & \shadecell{2} \\
BIC & \shadecell{0} & \shadecell{7} & \shadecell{0} \\
TRI & \shadecell{1} & \shadecell{1} & \shadecell{5} \\
\hline
\end{tabular}
\\[12pt]
\textbf{(c) p-value $\leq$ 0.50 (30 features)} & \textbf{(d) p-value $\leq$ 0.75 (47 features)} \\
\begin{tabular}{|c|ccc|}
\hline
\diagbox{True}{Pred} & DLT & BIC & TRI \\
\hline
DLT & \shadecell{5} & \shadecell{1} & \shadecell{1} \\
BIC & \shadecell{1} & \shadecell{6} & \shadecell{0} \\
TRI & \shadecell{2} & \shadecell{1} & \shadecell{4} \\
\hline
\end{tabular}
&
\begin{tabular}{|c|ccc|}
\hline
\diagbox{True}{Pred} & DLT & BIC & TRI \\
\hline
DLT & \shadecell{5} & \shadecell{1} & \shadecell{1} \\
BIC & \shadecell{1} & \shadecell{6} & \shadecell{0} \\
TRI & \shadecell{3} & \shadecell{2} & \shadecell{2} \\
\hline
\end{tabular}
\end{tabular}
\end{adjustbox}
\vspace{4pt}
\caption{Feature selection via statistical significance. \\
Confusion matrices using diagnosis features filtered by p-value thresholds ($\leq$0.10, $\leq$0.25, $\leq$0.50, $\leq$0.75). \\
Fatigue Rate is taken as 0.48, 0.72, 0.68 for DLT, BIC, TRI.}
\label{tab:confusion_pval_thresh}
\end{table}

\subsubsection{Simulator stepping durations}
Table~\ref{tab:sim_exp_comparison}(a) shows that musculoskeletal simulation with the default larger stepping duration (MyoSuite 20 ms, and MuJoCo 2 ms) significantly reduces the diagnosis performance. This happens because the large step sizes prevent convergence of the static optimization to the target sampled from the motion capture trajectory.

\subsubsection{Bell-shaped reference trajectories}
As mentioned in Section~\ref{sec:refernce_trajectory}, we use the real participants' motion capture trajectories as tracking references for the simulation. In this ablation study, we replaced them with theoretically generated minimum-jerk bell-shaped trajectories. Studies like~\cite{karniel1997model} have shown that humans employ such trajectories for reaching motions. As shown in Table~\ref{tab:sim_exp_comparison}(b), the diagnosis performance drops only by a small amount, validating the theoretical model.

\subsection{Naïve motion comparison}
This experiment is aimed at validating the impact of our proposed diagnostic features~\ref{sec:diagnostic_features}. Table~\ref{tab:sim_exp_comparison}(c) shows that naïve comparison of motion between the fatigued real participant and fatigued simulated model results in a significant drop in diagnosis performance. This is because normalization by the healthy motion trajectory, as done by the proposed diagnostic fatures, is important to account for the sim-to-real gap.

\begin{table}[ht!]
\centering
\renewcommand{\arraystretch}{1.2}
\setlength{\tabcolsep}{4pt}
\begin{adjustbox}{width=1\columnwidth}
\begin{tabular}{ccc}
\textbf{(a) MyoSuite @ 50 Hz} & \textbf{(b) Bell-shaped Ref. Trajectories} & \textbf{(c) Naïve motion comparison} \\
\begin{tabular}{|c|ccc|}
\hline
\diagbox{True}{Pred} & DLT & BIC & TRI \\
\hline
DLT & \shadecell{3} & \shadecell{4} & \shadecell{0} \\
BIC & \shadecell{1} & \shadecell{6} & \shadecell{0} \\
TRI & \shadecell{1} & \shadecell{3} & \shadecell{3} \\
\hline
\end{tabular}
&
\begin{tabular}{|c|ccc|}
\hline
\diagbox{True}{Pred} & DLT & BIC & TRI \\
\hline
DLT & \shadecell{5} & \shadecell{1} & \shadecell{1} \\
BIC & \shadecell{1} & \shadecell{4} & \shadecell{2} \\
TRI & \shadecell{3} & \shadecell{0} & \shadecell{4} \\
\hline
\end{tabular}
&
\begin{tabular}{|c|ccc|}
\hline
\diagbox{True}{Pred} & DLT & BIC & TRI \\
\hline
DLT & \shadecell{3} & \shadecell{1} & \shadecell{2} \\
BIC & \shadecell{1} & \shadecell{3} & \shadecell{3} \\
TRI & \shadecell{1} & \shadecell{2} & \shadecell{4} \\
\hline
\end{tabular}
\end{tabular}
\end{adjustbox}
\vspace{4pt}
\caption{Confusion matrices for (a) large simulator stepping durations, (b) bell-shaped reference trajectories, and (c) naïve motion comparison without use of our proposed diagnostic features.}
\vspace{-10pt}
\label{tab:sim_exp_comparison}
\end{table}

Overall, these experiments demonstrate that realistic simulation design and careful feature selection are both critical for reliable motion-based fatigue diagnosis.


\section{Discussion}

In summary, this study introduced a simulation-driven framework for analyzing upper-limb motion, demonstrating feasibility with around 70\% accuracy in identifying fatigue-specific muscle groups. Unlike prior fatigue assessments that typically focus on overall fatigue levels or rely on EMG signals, our approach highlights novelty by distinguishing which muscle group is fatigued using only kinematic features. This emphasizes the contribution of simulation-derived metrics in providing more granular insight into fatigue-related impairments.

At the control stage, simulations reproduced arm motion with reasonable fidelity, though some discrepancies remain due to limitations of PD-based torque estimation and static optimization. Exploring advanced control strategies such as model predictive control or reinforcement learning could improve smoothness, robustness, and adaptability of simulated motion.

The diagnostic framework itself also presents challenges, as diagnostic feature based classification currently relies on aggregated metrics across entire motion sequences, which may mask more localized or subtle impairment indicators. Future work should therefore investigate single-sequence or even intra-sequence diagnosis, which could enable earlier detection and finer resolution of classification. In addition, current healthy feature extraction depends mainly on processed joint angles; incorporating raw kinematic descriptors such as velocity profiles, inter-joint coordination, or trajectory synergies could enhance discriminative power.

Another limitation is the scope of validation. The framework at present serves primarily as a proof of concept to demonstrate that simulation-based feature generation can guide interpretation of experimental motion. Clinical translation will require testing with larger and more diverse subject groups, extending disorder-specific models beyond fatigue, and integrating multi-modal data sources such as EMG or IMU sensors alongside marker-based kinematics. Extension to non-recoverable musculoskeletal disorders will require better sim-to-real alignment because the healthy motion data will not be available for that case.

Finally, future work could explore subject-specific personalization of musculoskeletal models, adaptive calibration procedures for marker data, and integration with real-time feedback systems. These additions could increase robustness, improve diagnostic precision, and broaden applicability in both research and clinical contexts.

Overall, the results indicate a clear pathway for advancing contactless motion diagnosis: simulations can provide realistic impairment models, while feature extraction and validation against experimental data establish reliability. By refining both the simulation fidelity and feature design, the framework can evolve into a scalable, automated, and contactless diagnostic tool for HRI applications and early detection of musculoskeletal disorders.

\section*{Acknowledgment}
Financial and infrastructural support for this study was provided by the corresponding author’s institute through sponsored research awards and internal facility funding.

\printbibliography 
\end{document}